\documentclass[epsfig]{revtex4}
\usepackage{epsfig}
\usepackage{amsmath}
\usepackage{amstext}
\usepackage{amssymb}
\usepackage{amsfonts}
\usepackage{graphicx}
\usepackage{axodraw}
\usepackage{vmargin}
\usepackage{amscd}
\usepackage{bbold}
\usepackage{mathrsfs}
\usepackage{fancyheadings}
\usepackage{pifont}
\usepackage{pst-node}

\newcommand{\feyn}[1]{{#1}\!\!\!{\slash}}

\newcommand{\LL}{{\mathcal{L}}}

\newcommand{\qqbMM}[6]{
\begin{center}
\begin{picture}(500,130)(0,0)
\ArrowLine(80,35)(20,25) % q2
\Text(20,19)[l]{#2}
\ArrowLine(20,95)(80,85) % q1
\Text(20,100)[l]{#1}
\ArrowLine(145,23.5)(80,33.5)
\ArrowLine(80,36.5)(145,26.5)
\Text(145,17)[r]{#4}
\ArrowLine(80,86.5)(145,96.5)
\ArrowLine(145,93.5)(80,83.5)
\Text(145,102)[r]{#3}
\ArrowLine(80,85)(80,35)
\Vertex(80,35){3}
\Vertex(80,85){3}
\Text(75,60)[r]{#5}
\SetWidth{2}
\SetWidth{0.5}
\ArrowLine(220,60)(170,25)
\Text(170,19)[l]{#2}
\ArrowLine(170,95)(220,60)
\Text(170,100)[l]{#1}
\Vertex(220,60){3}
\ArrowLine(220,58)(270,58)
\ArrowLine(270,62)(220,62)
\Text(245,70)[c]{#6}
\Vertex(270,60){3}
\ArrowLine(320,25)(270,60)
\ArrowLine(270,60)(320,95)
\Vertex(320,25){3}
\ArrowLine(320,95)(320,25)
\Vertex(320,95){3}
\ArrowLine(320,23)(370,23)
\ArrowLine(370,27)(320,27)
\Text(330,15)[l]{#4}
\ArrowLine(370,93)(320,93)
\ArrowLine(320,97)(370,97)
\Text(330,85)[l]{#3}
\end{picture}
\end{center}}
\newcommand{\qqbqqb}[5]{	
\begin{minipage}{8.0cm}
\begin{picture}(340,130)(0,-10)
\ArrowLine(80,35)(20,25) % Diq
\Text(20,19)[l]{#2}
\ArrowLine(20,95)(80,85) % qb1
\Text(20,100)[l]{#1}
\ArrowLine(80,35)(80,85) % exchanged mesons
\ArrowLine(85,85)(85,35) % exchanged mesons
\ArrowLine(145,25)(86,35) % quarks
\Text(145,19)[r]{#4}
\ArrowLine(86,85)(145,90) % quarks
\Text(145,100)[r]{#3}
\Vertex(82.5,35){3}
\Vertex(82.5,85){3}
\Text(90,60)[l]{#5}
\SetWidth{2}
\SetWidth{0.5}
\ArrowLine(220,60)(170,25)
\Text(170,19)[l]{#2}
\ArrowLine(170,95)(220,60)
\Text(170,100)[l]{#1}
\ArrowLine(220,58)(270,58)
\ArrowLine(270,62)(220,62)
\Text(245,70)[c]{#5}
\ArrowLine(320,25)(270,60)
\ArrowLine(270,60)(320,95)
\Text(310,19)[l]{#4}
\Text(310,100)[l]{#3}
\Vertex(270,60){3}
\Vertex(220,60){3}
\end{picture}
\end{minipage}
}

\begin{document}
\title{Critical Opacity - a possible explanation of the fast thermalisation
times seen in RHIC experiments}
\author{F. Gastineau, E. Blanquier and J. Aichelin}
\address{
SUBATECH \\
Laboratoire de Physique Subatomique et des
Technologies Associ\'ees \\
University of Nantes - IN2P3/CNRS - Ecole des Mines de Nantes \\
4 rue Alfred Kastler, F-44072 Nantes, Cedex 03, France}

%\author{\begin{quote}
\begin{abstract}
The Nambu Jona-Lasinio Lagrangian offers an explication of the seemingly
contradictory observations that a) the energy loss in the entrance channel of
heavy ion reactions is not sufficient to thermalize the system and that b) the 
observed hadron cross sections are in almost perfect agreement with 
hydrodynamical calculations. According to this scenario, a critical opacity 
develops close to the chiral phase transition which equilibrates and hadronizes 
the expanding system very effectively. It creates as well radial flow and, if the system is
not isotropic, finite $v_2$ values. 
\end{abstract}
%\end{quote}}
\date{\today}
\maketitle
%\setpapersize{A4}
      %    Width of rule between columns.
\section{Introduction}
Some of the most surprising results of the ongoing RHIC experiments are the
large radial flow $\beta_{rad} \approx 0.5$, the large $v_2$  values, where $v_2$ is the second Fourier coefficient of a Fourier
expansion in $\phi_p$ of $dN/(dyp_tdp_t d\phi_p)$, and the fact that hydrodynamical
models describe not only the total but even the differential 
anisotropy $v_2(p_t)$ up to a transverse
momentum of $p_t$ = 2 GeV \cite{h1,h2,h3,h4,h5}. The initial energy densities,
at $\tau = 1fm/c$, employed in these calculation are between 4 and 6 $ GeV/fm^3$ 
and agree quite well with the Bjorken estimate \cite{adc}. This seems to
indicate that the system is at that time close to local equilibrium.
For a survey of the results of the hydrodynamical approaches
see \cite{huo}.
The problem with these findings is that one does not understand how the system comes
so fast to equilibrium. The energy loss of incoming partons is of the order 
of a 1 GeV/fm \cite{tho,gyu}. Independent of the reaction scenario employed 
(minijets\cite{mjet} or Weizsaecker-William fields of the incoming nuclei 
which materialize
into on-shell gluons \cite{mul}) the time needed for equilibration is much 
longer. For a recent review see \cite{ser}.
 
All these models share the feature that at energies below the validity of pQCD
phenomenological approaches have to be employed. Therefore it may be 
worthwhile to use another phenomenological model which has
been shown to be quite successful in regions where it can be compared with
QCD, the Nambu Jona Lasinio (NJL) model\cite{ebert,kle92}. Supplemented by a 
t'Hooft determinant the NJL Lagrangian has the same symmetries as the QCD 
Lagrangian which are known to be the basis of many properties of quark and
hadronic matter. It predicts\cite{gas}, as pQCD
\cite{Pis,Wil} calculations, color
flavour locking at low temperatures and high densities, allows to describe the
meson masses at low density and low temperature \cite{kle92} and provides a simple approach
to study the chiral phase transition. It predicts as well the tricritical 
point in the $\rho - T$ phase diagram which has recently been predicted by 
lattice QCD calculations\cite{fod}. 
Of course this model has as well its deficiencies: Based on a local 4 - Fermion
interaction it is not renormalizable and therefore a cutoff $\Lambda$ has to 
be employed to regularize the loops. Furthermore due to the locality of the
interaction confinement is not present and gluons do not appear as degrees of
freedom. Due to these drawbacks it is hard to judge the quantitative prediction
of this model but it can certainly serve for qualitative studies.

Using this model we will show that due to the interplay between the 
quark and meson masses close to the chiral phase transition the system may 
develop a critical opacity where all  s - channel transition rates become very
large. This it true for the elastic as well as for the hadronization  cross 
sections. These large cross sections equilibrate very efficiently the expanding 
plasma and create a gaz of mesons although confinement is 
absent in the NJL Lagrangian. Since the equilibration takes place during the
expansion a strong radial flow develops which is seen the data and which is as
well not understood yet.
  
\section{The NJL approach}
The NJL model is the simplest low energy approximation 
of QCD. It describes the interaction between two quark currents as a 
point-like exchange of a perturbative gluon \cite{ebert}. 
This local interaction is given by
\begin{equation}
\LL^{int} =  \kappa \sum ^{N_c^2-1}_{c=1}\sum_{i,j}^3(\bar{q}_{i,\alpha }[\gamma
_{\mu}\lambda^c]_{\alpha \delta}q_{i,\delta})
(\bar{q}_{j,\gamma }[\gamma ^{\mu}\lambda^c]_{\gamma \beta}q_{j,\beta}).
\end{equation}
where we have explicitly shown the flavor (i,j) and color/Dirac 
($\alpha,\beta,\gamma,\delta$) indices. We normalize $\sum_{i=0}^8 \lambda^i_{\alpha \beta}
\lambda^i_{\beta \alpha} = 2$. Applying a Fierz transformation in color space to this interaction
the Lagrangian separates into two pieces: an attractive color singlet interaction between a quark and an
antiquark (\( \LL _{(\bar{q}q)} \)) and a repulsive color anti-triplet interaction between
two quarks \( \LL _{(qq)} \) which disappears in the large $N_c$ limit. 
Here we are only interested in the color singlet channel (the color octet
channel gives diquarks and can be used to study baryons \cite{gas}):
\begin{equation}
\LL =\LL _{0}+\LL _{(\bar{q}q)}+\LL _{A}
\end{equation}
where \( \LL _{0} \) is the free kinetic part.
Concentrating on the dominant scalar and pseudo scalar part in Dirac space 
we find the following Lagrangian which we use here:
\begin{eqnarray}
\LL  & = & \sum _{f=\{u,d,s\}}\left[\bar{q}_{f}(i\feyn 
{\partial }-m_{0f})q_{f}+G_{S}\sum ^{8}_{a=0}\left[ (\bar{q}_{f}\lambda _{F}^{a}q_{f})^{2}+(\bar{q}_{f}i\gamma _{5}\lambda
_{F}^{a}q_{f})^{2}\right]\right] 
\label{Lagr_allgemein} \\
 &  & +G_{D}\left[ det [\bar{q}_{f}(1-i\gamma _{5})q_{f}] + det [\bar{q}_{f}(1+i\gamma _{5})q_{f}]\right] .\nonumber 
\end{eqnarray}
The first term is the free kinetic part, including the flavor dependent current
quark masses \( m_{0f} \) which break explicitly the chiral symmetry of the
Lagrangian. The second part is the scalar/pseudoscalar interaction in the mesonic
channel. It is diagonal in color. We have added the six point interaction 
in the form of the t'Hooft determinant which breaks explicitly the \( U_{A}(1) \)
symmetry of the Lagrangian. The \( det \) runs over the flavor degrees of freedom,
consequently the flavors become connected.

The model contains five parameters: the current mass of the light and strange
quarks, the coupling constants \( G_{D} \) and \( G_{S} \) and the momentum
cut-off \( \Lambda  \),  are fixed by physical observables: the pion and
kaon mass, the pion decay constant, the scalar quark condensate $<\bar qq>$ and the mass difference between \( \eta  \)
and \( \eta ' \). We will employ the parameters set:$m^0_{q}= 4.75\ MeV,
m^0_{s}= 147 \ MeV , G_S/\Lambda^2= 1.922, G_D/\Lambda^5 = 10,
\Lambda = 708 \ MeV$

The temperature and density dependent masses of the quarks are obtained by
reducing the above Lagrangian to a one particle Lagrangian by contracting the
remaining field operators in all possible ways. The meson masses
are obtained by solving the Bethe Salpeter equation in the $q\bar q$ channel.
The details of both calculations are found in ref. \cite{kle92}.

\begin{figure}[hbt]
\includegraphics[width=8.5cm,angle=270]{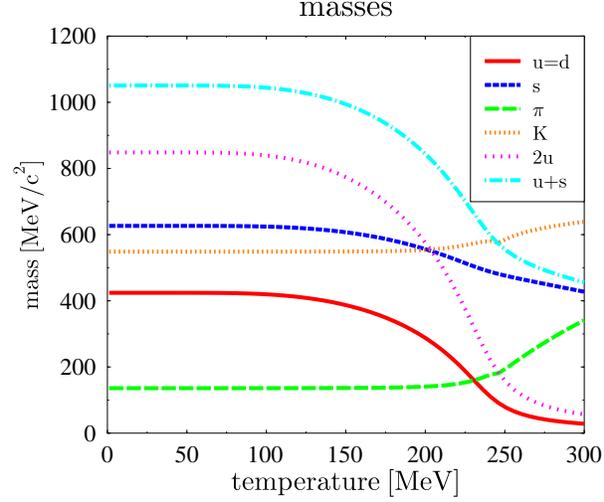}
\caption{Masses of the the pseudo scalar mesons and of the quarks in the NJL
approach.}
\label{mass}
\end{figure}

With these parameter we obtain the meson and quark masses displayed in
fig.\ref{mass}.

\section{Cross sections}
If created in heavy ion collisions the quark gluon plasma will expand rapidly. 
Therefore, not the static properties of the
theory but the cross sections between constituents become dominant. In the
NJL model these cross sections can be calculated via a $1 /N_c$ 
expansion \cite{huef,Reh96}. 

\begin{figure}[hbt]
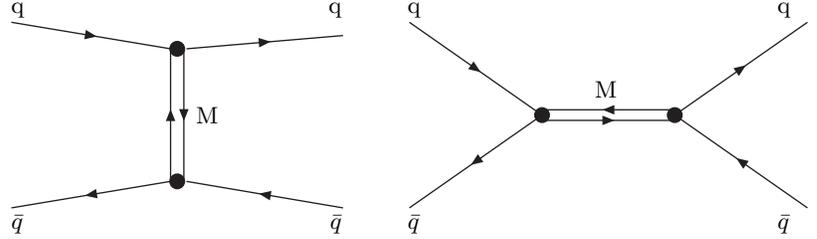

%\begin{center}
\hspace*{2cm}\qqbqqb{q}{$\bar{q}$}{q}{$\bar{q}$}{M}{\;}\vspace*{1cm}
%\end{center}
\caption{Feynman diagrams of $q+\bar{q} \rightarrow q+ \bar{q} $.}
\label{qqbqqb}
\end{figure}

\begin{figure}[hbt]
%\vspace*{-6cm}
\includegraphics[width=8.5cm]{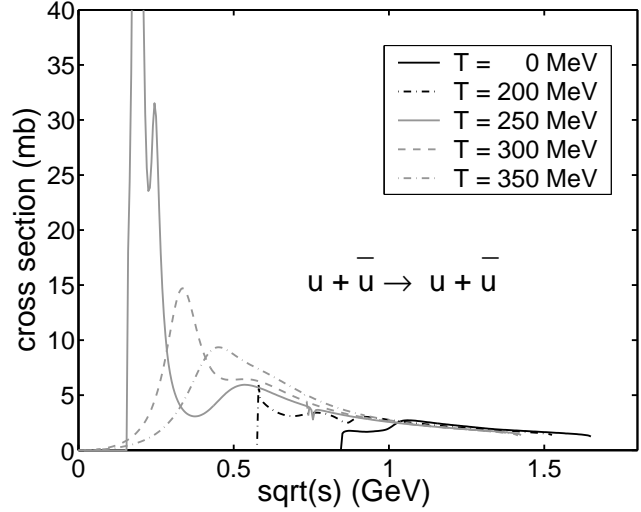}
\caption{Elastic cross section $\sigma_{u\bar u\rightarrow u\bar u.}$ for
different temperatures as a function of $\sqrt{s}$.}
\label{sela}
\end{figure}

We start out with the elastic cross sections. The $qq\rightarrow qq$ cross
section has no s-channel contribution and is of the order of some mb.
It is not important for the phenomena discussed in this letter.
The Feynman diagrams for the $q\bar{q}\rightarrow q\bar{q}$ cross section is shown in fig.\ref{qqbqqb}. 
The matrix elements in the t and s channel are given by
\begin{eqnarray}
-i{\mathcal{M}}_t & = &
\delta_{c_1,c_3}\delta_{c_2,c_4}\bar{u}(p_3)Tu(p_1)[i{\mathcal{D}}^S_t(p_1-p_3)]{v}(p_4)T\bar{v}(p_2) \\
& + &
\delta_{c_1,c_3}\delta_{c_2,c_4}\bar{u}(p_3)(i\gamma_5T)u(p_1)[i{\mathcal{D}}^P_t(p_1-p_3)]v(p_4)(i\gamma_5T)\bar{v}(p_2) \nonumber \\
-i{\mathcal{M}}_s & = &
\delta_{c_1,c_2}\delta_{c_3,c_4}v(p_2)Tu(p_1)[i{\mathcal{D}}^S_s(p_1+p_2)]{v}(p_4)T\bar{u}(p_3) \\
& + & \delta_{c_1,c2}\delta_{c_3,c_4}\bar{v}(p_2)(i\gamma_5T)u(p_1)[i{\mathcal{D}}^P_u(p_1+p_2)]{v}(p_4)(i\gamma_5T)\bar{u}(p_3) \nonumber \; ,
\end{eqnarray}
where $p_1(p_2)$ is the momentum of the incoming $q(\bar q$) and 
$p_3 (p_4)$ that from the outgoing $q(\bar q$). The $c_i$ are
the color indices and T are the isospin projections on the mesons. ${\mathcal{D}}^S$ and ${\mathcal{D}}^P$ are the meson propagators of the form 
\begin{equation}
{\mathcal{D}}^{S/P} = \frac{2G_S}{ 1 - 2G_S\Pi^{S/P}}
\label{res}
\end{equation}
with $\Pi^{S/P}$ being the polarization tensor in the scalar/pseudoscalar 
channel. This cross section is displayed in fig. \ref{sela}.
We observe, as expected, an elastic cross section of the order of several 
millibarn but close to the critical temperature ($T_c = 240 MeV$)
the cross section increases dramatically due to the resonance structure in the s-channel.

This resonance in the s-channel dominates also the hadronization matrix elements close to the threshold whose Feynman diagrams are given
in fig. \ref{qqbmm} 
\begin{figure}[hbt]
\begin{center}
\qqbMM{q}{$\bar{q}$}{M}{M}{q}{M}
\end{center}
\caption{Generic form of Feynman diagrams for t and s channel of  $q+\bar{q} \rightarrow M+ M $. The u channel is obtained by exchanging
the mesons of the t channel.}
\label{qqbmm}
\end{figure}
%%%       hier die anderen Bilder hin, damit sie dahinter kommen

\begin{figure}[hbt]
\begin{center}
\includegraphics[width=7.5cm]{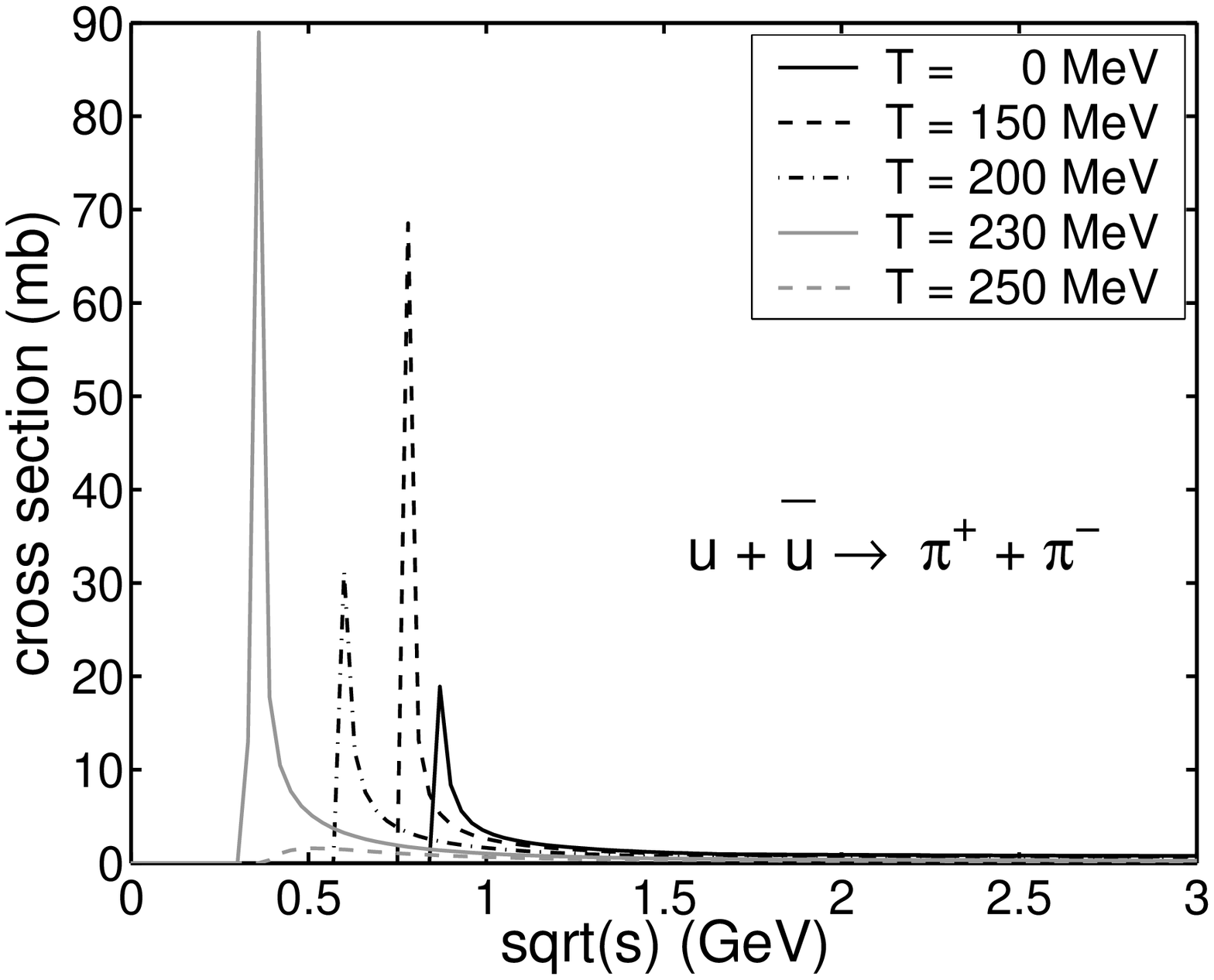}
\includegraphics[width=7.5cm]{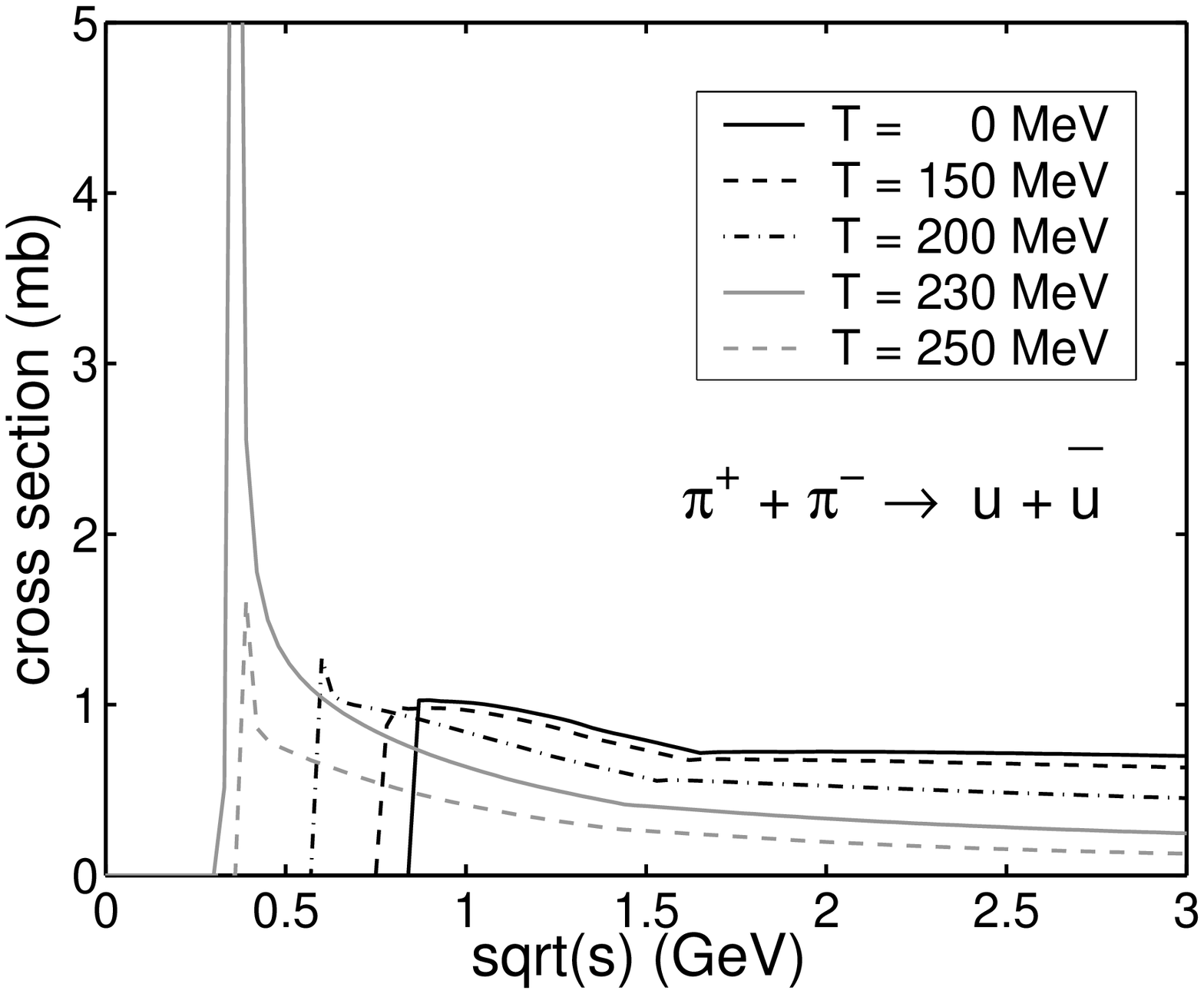}
\end{center}
\caption{Cross section for $u+ \bar{u} \rightarrow \pi^+ + \pi^-$ and for the inverse reaction as a function of $\sqrt{s}$ for different
temperatures.}
\label{sine}
\end{figure}

\begin{figure}[hbt]
\begin{center}
\includegraphics[width=7.5cm]{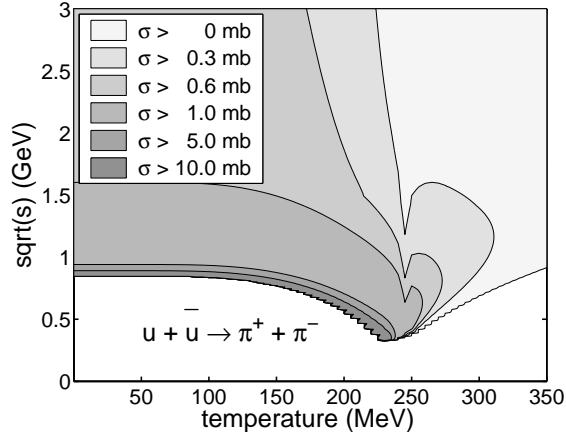}
\end{center}
\caption{Size of the cross section for  $u+ \bar{u} \rightarrow \pi^+ + \pi^-$ in the $\sqrt{s}$ - temperature plane.}
\label{plants}
\end{figure}
\begin{figure}
\centering
\begin{minipage}{7cm}
  \includegraphics[width=7cm]{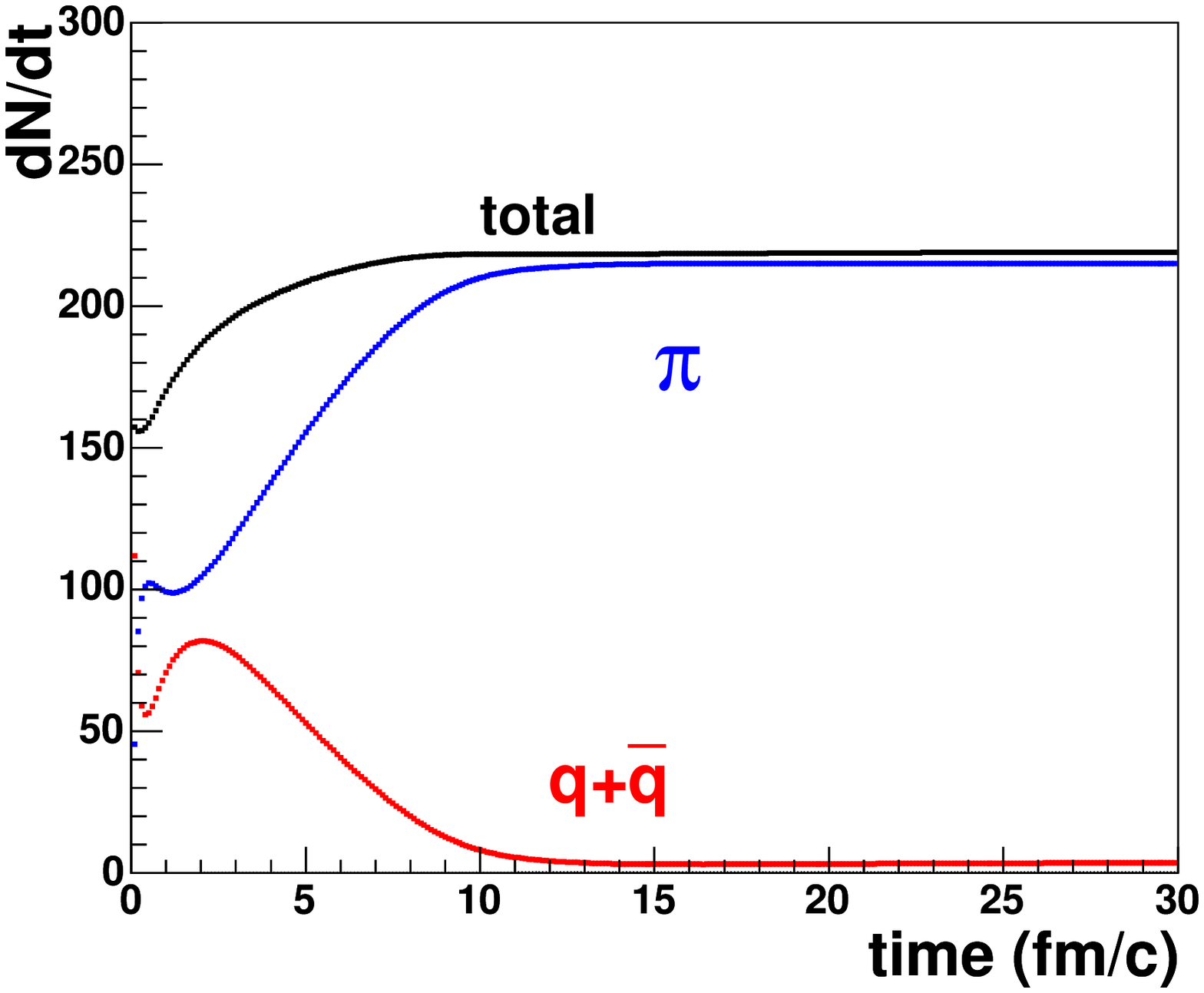}
  \caption{Number of pions and quarks as a function of time.}
\label{hadro:nbqpi}
\end{minipage}
\begin{minipage}{7cm}
  \includegraphics[width=7cm]{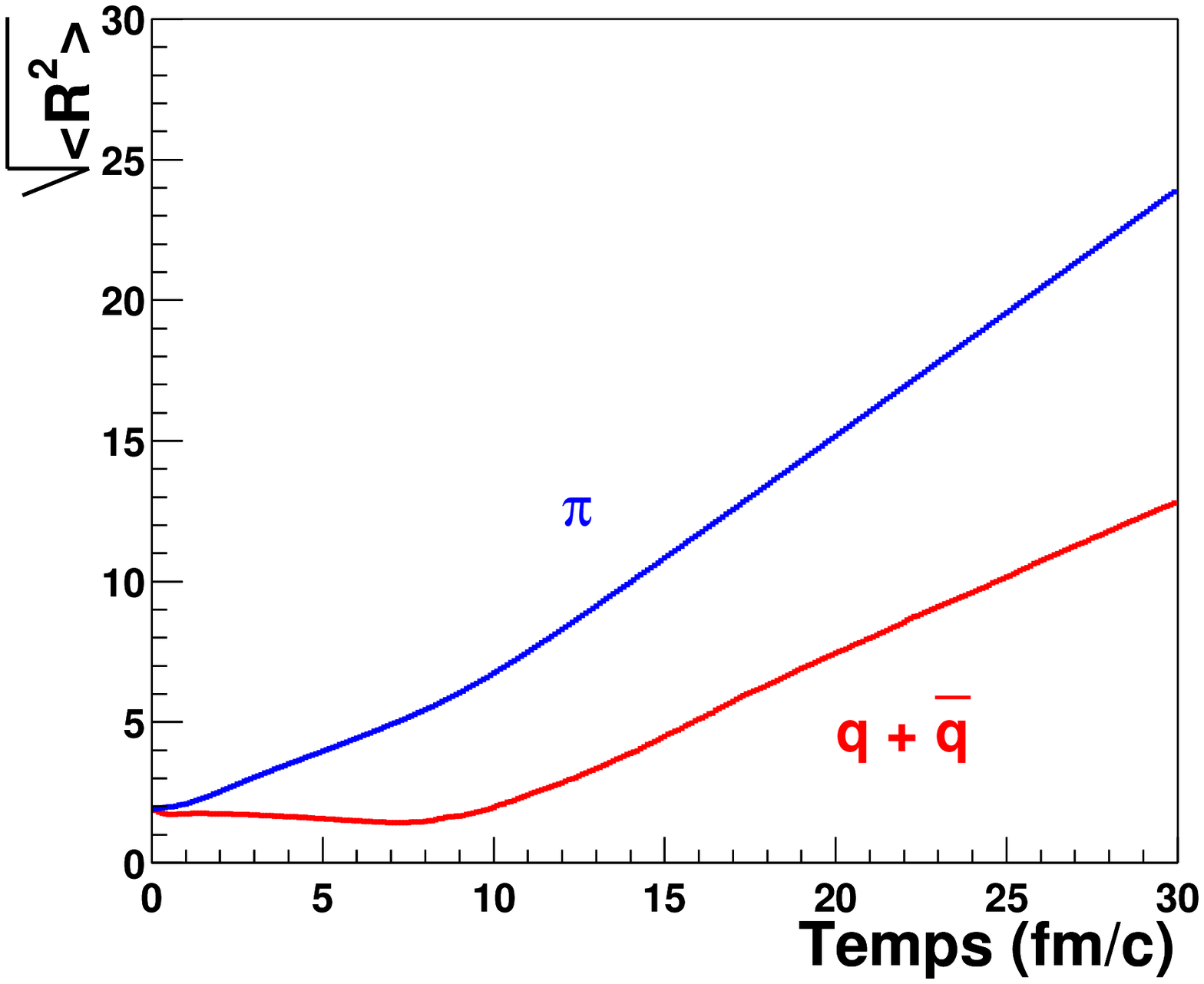}
  \caption{Root mean square radius of the different species as a function of time.}
  \label{hadro:rad}
\end{minipage}
\end{figure}

with the matrix elements:
\begin{eqnarray}
-i{\mathcal{M}}_t &=& f_t \delta_{c_1 c_2} \bar{v}(p_2) i \gamma_5 i g_1 S_F(p_1-p_3) i \gamma_5 ig_2 u(p_1) \\
-i{\mathcal{M}}_u &=& f_u \delta_{c_1 c_2} \bar{v}(p_2) i \gamma_5 i g_1 S_F(p_1-p_4) i \gamma_5 ig_2 u(p_1) \\
-i{\mathcal{M}}_s &=& \bar{v}(p_2) \delta_{c_1c_2} f_s i {\mathcal{D}}(p_1+p_2) \Gamma(p_1+p_2;p_3) ig_1 ig_2u(p_1). 
\end{eqnarray}
We discuss here as an example  the hadronization cross section for $u+ \bar{u} \rightarrow \pi^+ + \pi^-$ 
(which has two s and one t channel) and the cross section for the inverse reaction which are displayed in fig. \ref{sine}.  

Around $T_c$ the transition amplitude diverges close to the threshold, therefore the cross sections in both directions have 
their maximum there. The difference comes from the different flux and phase space factors. 
\section{Expanding Plasma}
These cross sections suggest the following reaction scenario: as soon as projectile and target nuclei overlap the
density is that high that the incoming nucleons overlap and partons can travel in this matter. By pQCD cross sections, which are of the
order of some mb, the partons scatter but these cross sections are not strong enough to equilibrate the system. While the system expands 
the local available center of mass energy and hence the local temperature lowers. The closer the system comes
to the phase transition the larger becomes the elastic cross sections. Due to this critical opacity the system behaves 
close to the phase transition more like a liquid then like a plasma. 
In this phase the system approaches thermal equilibrium. The large cross sections during the expansion create in addition
a large radial flow and - if the system is azimuthally not isotropic - also $v_2$ values. When passing the critical temperature
the large hadronization cross sections become effective because the created mesons do not decay anymore.  The inverse cross section 
is small due to the kinematic conditions. The system cools locally and the maximum of the cross section moves to larger
values of $\sqrt{s}$. This means that quark pairs with larger center of mass energies can now hadronize. During the expansion
the local temperature passes from  $T > T_c$ to  0. Therefore the cross sections are large for a large interval of 
$\sqrt{s}$ as seen in fig. \ref{plants}.  

There remains the quantitative question whether the strength of these cross sections is sufficient to hadronize the system completely:    
In order to check this we perform calculations using the Quantum Molecular Dynamics approach \cite{aic} which has been
successful used to describe heavy ion reactions in this energy domain. In this approach partons are presented as Gaussians 
whose centers in momentum and coordinate space follow Hamilton's equations. In addition they interact via cross sections calculated from
the same Hamiltonian. Fig.\ref{hadro:nbqpi} presents a model calculation for 30 quarks and 30 antiquarks and 100 pions
initialized with isotropic distributions in coordinate and momentum space for an energy density of $3 GeV/fm^3$ which give a temperature
slightly above $T_c$. 
There pions are unstable and therefore the mesons can decay into the lighter quarks.
When the system approaches the phase transition the quark condensate $<q\bar{q}>$ increases: the quarks become heavier and this 
energy is taken out from the relative motion.
At the phase transition the mesons become stable and lighter than the quarks. They escape from the system in which the quarks remain.
By emitting pions the remaining quark system cools down until finally only 1 quarks and 1 antiquark are left which have not found 
a partner to created two pions. Thus, despite of the fact that confinement is not enforced by the Lagrangian, under the condition 
of an expanding plasma confinement is almost complete.

In conclusion we have shown, that the thermalisation observed in the spectra in relativistic heavy ion reactions may be reconciled 
with the small pQCD cross section assuming that the system passes during its expansion a phase transition with critical opalescence
as predicted by the NJL - Lagrangian. Both, elastic as well as inelastic cross section, become large over a large kinematic region
which is covered by the expansing system and lead to a local thermalisation as soon as the system comes to the phase transition.
Because this transition takes places in an already expanding system, radial flow is created as observed experimentally. The observed $v_2$
values are an image of the asymmetry at the beginning of the expansion.

\end{document}